\documentclass[final]{svjour2}
\usepackage{tikz}
\usepackage{graphicx}
\usepackage{rotating}
\usepackage{amssymb}
\usepackage{mathptmx}
\usepackage[numbers]{natbib}
\makeatletter
\journalname{Journal of Low Temperature Physics}

\bibpunct{}{}{,}{s}{}{,}

\begin{document}

\newcommand{\hdblarrow}{H\makebox[0.9ex][l]{$\downdownarrows$}-}
\title{Defect formation in superconducting rings: external fields and finite-size effects}

\author{D.J.~Weir \and R.~Monaco \and R.J.~Rivers}
\institute{
D.J.~Weir
  \at  Helsinki Institute of Physics, Gustaf H\"{a}llstr\"{o}min katu 2a, 00014 Helsinki, Finland\\\email{david.weir@helsinki.fi}
\and
R.~Monaco
 \at Istituto di Cibernetica del CNR, Comprensorio Olivetti, 80078 Pozzuoli (Na), Italy
\and
R.J.~Rivers
  \at Theoretical Physics Group, Blackett Laboratory, Imperial College London, SW7 2AZ, UK
}

\date{August 15, 2012}

\maketitle


\begin{abstract}

Consistent with the predictions of Kibble and Zurek, scaling behaviour
has been seen in the production of fluxoids during temperature
quenches of superconducting rings. However, deviations from the canonical behaviour arise because of finite-size effects and stray external fields. 

Technical developments, including laser heating and the use of long Josephson tunnel junctions, have improved the quality of data that can be obtained. With new experiments in mind we perform large-scale 3D simulations of quenches of small, thin rings of various geometries with fully dynamical electromagnetic fields, at nonzero externally applied magnetic flux. We find that the outcomes are, in practice, indistinguishable from those of much simpler Gaussian analytical approximations in which the rings are treated as one-dimensional systems and the magnetic field fluctuation-free.

\end{abstract}

\section{Introduction}

Many systems of interest in condensed matter physics exhibit a second-order phase transition that can be studied in the laboratory. For such transitions one would expect, under adiabatic change, that the correlation length of the system could grow to be as large as the system size. The system therefore orders itself simultaneously and in the same manner everywhere. In practice, however, causality (the finite speed at which order can propagate) restricts the size of the domains within which the system can order itself and the correlation lengths are rather smaller. The faster the transition, the shorter the largest correlation length that is reached, and hence the smaller the domains.

If the phase transition in question results in a spontaneously broken symmetry, different choices of the order parameter can be made by domains that have been causally disconnected while the system was out of equilibrium. If there is frustration then topological defects will form, which may persist long enough to be detected.

Such spontaneous formation of topological defects from phase transitions out of equilibrium is frequently termed the Kibble-Zurek (KZ) scenario. The process outlined above was discussed by Kibble\cite{Kibble:1976sj,Kibble:1980mv} in the context of Grand Unified Theories (GUTs) describing the physics of the early universe. Zurek independently suggested that the same process could occur in condensed matter systems\cite{Zurek:1985qw,Zurek:1993ek}.

The most straightforward (and elegant) manifestation of the KZ scenario involves a linear temperature quench from one phase to another. Let us parameterise this linear quench by the time taken $\tau_\mathrm{Q} = T_\mathrm{c} (dT/dt)^{-1}_{T_\mathrm{c}}$, where $T_\mathrm{c}$ is the critical temperature of the second-order phase transition in question.

If the system is initially homogeneous and isotropic it is proposed that the initial defect separation $\bar{\xi}$ will be given by a scaling form
\begin{equation}
\label{eq:kzlaw}
\bar{\xi} \approx \xi_0 \left(\frac{\tau_\mathrm{Q}}{\tau_0}\right)^\sigma,
\end{equation}
where $\xi_0$ and $\tau_0$ are system-dependent. The exponent $\sigma$, however, can be derived from critical exponents valid for the adiabatic transition and so the essential university of the phase transition can be seen even in this out-of-equilibrium behaviour.

Unfortunately, we have yet to see evidence for defects in cosmological observations, although they are predicted in all reasonable models that incorporate unification. However, Eq.~(\ref{eq:kzlaw}) has been tested in a wide variety of different condensed matter systems. Restricting our attention here to experiments involving type-II superconductors, it has been demonstrated\cite{Polturak,2006PRL} that such arguments are applicable to Josephson tunnel junctions and to isolated superconducting rings\cite{2009PRB}; these experimental results are supported by simulations\cite{Yates:1998kx,Donaire:2004gp}.

There is one significant difference between cosmological and condensed matter systems; whereas cosmological phase transitions take place in effectively infinite volume, those in condensed matter experiments are limited to finite -- and often quite small -- sizes. This can lead to edge effects, and allowances should be made for the different physics close to the boundaries of such systems.

We shall concentrate here on fluxoid formation experiments in rings of Niobium. If the ring is sufficiently narrow, then it is plausible that the ring behaves like a one-dimensional object with finite size and periodicity. This accords with the KZ picture, for which Eq.~(\ref{eq:kzlaw}) is now understood as a perimeter law. A previous study tested these assumptions and examined how the KZ scenario unfolds for slow quenches in small 1D systems with periodicity\cite{WeirRivers}. 

Here, this work is extended in two principal ways.

First, we extend our simulations from one dimension to three dimensions, to describing a flat annulus of superconductor in three-dimensional space, in which the electromagnetic field is fully dynamical (impossible in one or two-dimensional simulations). This allows us to test different ring geometries (the KZ picture implies only perimeter scaling behaviour, indifferent to ring shape and width), and the breakdown of the KZ scaling laws for small systems. In addition, with dynamic electromagnetic fields, we can determine the likelihood of observing Hindmarsh-Rajantie magnetic field freeze-out\cite{Hindmarsh:2000kd}, which also goes beyond the KZ scenario.

Secondly, we discuss the effect of an external magnetic field on the superconducting phase transition and compare with simple theoretical estimates for how this biases the fluxoid trapping rate. Experimentally, it is impossible to eliminate stray magnetic fields completely; if we can understand the dependence of fluxoid production on the applied field than we can reduce errors due to stray field bias.

\section{Analytic approximations}

As we say, realistic systems can give deviations from the simple behaviour of Eq.~(\ref{eq:kzlaw}) because of finite size and stray fields. To provide an estimate of the likelihood of finding spontaneously produced fluxoids within our ring taking these into account, against which we can test our simulations, we adopt simple Gaussian ansatze, assuming that we can treat the annulus as a 1D system.

Let us denote the winding number of the complex order parameter field $\Phi$ (which mimics Cooper pairs) along the ring by $n$. That is, if the ring has circumference $C$, the total winding number is
\begin{equation}
n = \frac{1}{2\pi} \oint d\ell \; \frac{\partial \mathrm{Arg}\,
  \Phi}{\partial \ell} = \frac{ \theta}{2\pi}
\end{equation}
where $ \theta$ is the change in phase around the ring. The probability of seeing $n$ fluxoids or anti-fluxoids is $f_n = p_n + p_{-n}$. We shall concentrate on the slow quench regime in which $f_1 \ll f_2$ and can ignore the production of more than one unit of flux.

\subsection{Finite size effects}

We have discussed this elsewhere\cite{WeirRivers} but, in brief, we assume that the winding number density is a Gaussian random variable. If $C$ is the circumference of the ring the relevant quantity is ${\bar\xi}/C$. This determines whether the ring is `large' or `small' according as the ratio is small or large. Finite size effects arise either because the ring is physically small or because the quench is slow. 

In the absence of an external field this Gaussian ansatz is sufficient to reproduce the KZ scenario for large rings, with exponent $\sigma = 0.25$. However, for small rings we find exponential damping, in which Eq.~(\ref{eq:kzlaw}) is replaced by
\begin{eqnarray}
\label{eq:expdamping}
\ln f_1 &\approx& - A\bar{\xi}^2/C^2 + {\mbox const.}
\end{eqnarray}
The coefficient $A$ is not simply estimated. Eq.~(\ref{eq:expdamping}) is in contrast to the proposal that the slope be doubled\cite{2009PRB}, which may arise for a disc with a small central hole, but which we have not tested here.

\subsection{External field effects}

 Now make the slightly different assumption that $\mathrm{Arg}\; \Phi$ itself has Gaussian correlations around the ring.  In the presence of a field, it is proposed that the accumulated phase change $\theta$ then has a Gaussian probability distribution $G(\theta; \phi_\mathrm{f})$
\begin{equation}
G(\theta; \phi_\mathrm{f}) = \frac{1}{\sqrt{2\pi \sigma^2}} \exp -\frac{(    \theta - 2\pi\phi_\mathrm{f})^2}{2 \sigma^2}
\end{equation}
where we have assumed that the mean is biased from zero by the applied normalised flux $\phi_\mathrm{f}$ through the ring; the deviation $\sigma$ depends on geometry and quench time. With such a distribution, the probability $p_n$ of ending up with a given winding number $n$ is given by
\begin{equation}
p_n(\phi_\mathrm{f}) = \int_{-\pi+2n\pi}^{\pi + 2n\pi} d\theta \; G(\theta;
\phi_\mathrm{f}).
\end{equation}
In the presence of a stray residual field, $\phi_\mathrm{r}$, we replace $\phi_\mathrm{f}$ by $\phi_\mathrm{f}-\phi_\mathrm{r}$. The probability of trapping $n$ fluxoids can be found to be
\begin{equation}
\label{eq:f0}
p_n(\phi_\mathrm{f}) = \frac{1}{2}\left[\mathrm{erf}\,
  \frac{\phi_\mathrm{f} - \phi_\mathrm{r} - n + 0.5}{s} - \mathrm{erf}\,\frac{\phi_\mathrm{f} -
    \phi_\mathrm{r} - n -0.5}{s}\right],
\end{equation}
where the dependence on quench time and geometry is parameterised by $s$. Similar expressions can be derived for the probability of trapping other numbers of fluxoids. This calculation unfortunately does not allow us to determine the dependence of $s$ on $\tau_\mathrm{Q}$. Nevertheless, we can fit data for different applied magnetic fields at fixed $\tau_\mathrm{Q}$ to Eq.~(\ref{eq:f0}) with only $s$ and $\phi_\mathrm{r}$ as free parameters to yield $f_0$. We will test the validity of this ansatz (with $\phi_\mathrm{r}=0$) in our simulations.

\section{Simulations}
\label{sec:simulations}

\begin{figure}
\begin{center}
\begin{tikzpicture}[x=0.65\linewidth,y=0.65\linewidth]

\draw (0.2,0.5) rectangle (0.8,0.9);
\fill[lightgray] (0.2625,0.5625) rectangle (0.7375,0.8375);
\fill[white] (0.325,0.625) rectangle (0.675,0.775);

\draw(0.7375,0.8375) node[below left] {$\mathcal{R}$};

\draw[<->] (0.2,0.475) -- (0.8,0.475);
\draw(0.5,0.475) node[below] {$L_x$};
\draw[<->] (0.175,0.5) -- (0.175,0.9);
\draw(0.175,0.7) node[left] {$L_y$};

\draw[<->] (0.325,0.65) -- (0.675,0.65);
\draw(0.675,0.65) node[above left] {$\Delta x$};
\draw[<->] (0.35,0.625) -- (0.35,0.775);
\draw(0.35,0.7775) node[below right] {$\Delta y$};

\fill[lightgray] (0.1225,0.025) -- (0.5975,0.025) -- (0.7775,0.1) -- (0.3025,0.1)--(0.1225,0.025);
\fill[white] (0.2258,0.042) -- (0.5758,0.042) -- (0.6766,0.084) -- (0.3266,0.085)--(0.2258,0.042);

\draw (0,0) -- (0.6,0) -- (0.9,0.125) -- (0.3,0.125) -- (0,0);
\draw (0,0.2) -- (0.6,0.2) -- (0.9,0.3) -- (0.3,0.3) -- (0,0.2);
\draw (0,0) -- (0,0.2);
\draw (0.6,0) -- (0.6,0.2);
\draw (0.9,0.125) -- (0.9,0.3);
\draw (0.3,0.125) -- (0.3,0.3);

\draw[->] (0.95,0.125) -- (0.95,0.3);
\draw(0.95,0.2125) node[right] {$B_\mathrm{ext}$};

\draw[<->] (-0.025,0) -- (-0.025,0.2);
\draw(-0.025,0.1) node[left] {$L_z$};

\end{tikzpicture}
\end{center}
\caption{Simulation setup. In lattice units (we typically take $a=0.5$), the ring -- the shaded region marked $\mathcal{R}$ -- is always of a constant width of 5. The border around the ring is also 5 units thick; the size $\Delta x\times \Delta y$ of the
  central void therefore determines $L_x$ and $L_y$; we take $L_z =
  5$. The external magnetic field $B_\mathrm{ext}$ is applied in the
  direction shown. Periodic boundary conditions are used in all three directions.}
\label{fig:setup}
\end{figure}
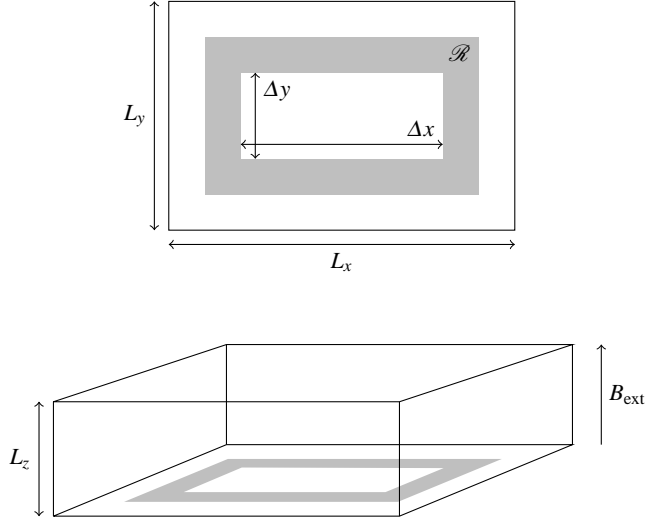

To test the approximations outlined in the previous section, and investigate the extent to which the physics of a superconducting ring is one-dimensional, we carried out simulations. In our system, illustrated in Fig.~\ref{fig:setup}, the superconducting ring (in which the complex scalar field $\Phi$ is defined) is surrounded on all sides by a box in which the standard $\mathrm{U}(1)$ vacuum electrodynamics is simulated. Although the superconductor is thereby taken to be a thin planar film, the three-dimensional simulation allows for any nontrivial correlations of the magnetic field to play a role in addition to the KZ mechanism. We can therefore study both mechanisms that may be thought to affect fluxoid formation in this system.

We use a gauge-invariant Langevin formulation for gauge and scalar fields, making use of Krasnitz's work on formulating Langevin equations for systems with first-class constraints\cite{Krasnitz:1995xi}.

A noncompact formulation of the $\mathrm{U}(1)$ gauge field is used, so that it can take any value on each link rather than being defined up to $2\pi$. The Hamiltonian for the system is
\begin{eqnarray}
H &=& \frac{1}{2} \sum_{x,i} a^3 \left[E_i(x)^2 + \sum_{jk}
\left(\epsilon_{ijk} \Delta^+_j A_k(x)\right)^2 \right] - 2 \sum_{x \in\mathcal{R},i} \mathrm{Re} \; \Phi^* (x) U_i(x) \Phi(x+i)  \nonumber \\
&& + a^2 \sum_{x\in\mathcal{R}}
 \left[\Pi^* \Pi + \left(m^2 +
\frac{4}{a^2}\right) \Phi^* \Phi + \frac{\lambda}{2} \left(
\Phi^* \Phi \right)^2 \right]
\end{eqnarray}
where $U_i(x) =e^{iaeA_i(x)}$, and the finite differences are
\begin{equation}
\Delta^+_i f(x) = \frac{f(x + a\hat{\imath})-f(x)}{a}; \qquad \Delta^-_i f(x) = \frac{f(x) - f(x - a\hat{\imath})}{a}.
\end{equation}
Note that $\Phi$ is only defined on a 2-dimensional superconducting ring region denoted $\mathcal{R}$. The only gauge fixing that has been carried out is $A_0=0$; `temporal gauge'. This yields the Gauss constraint
\begin{equation}
\sum_i \Delta^- E_i = \frac{2e}{a} \mathrm{Im}\; \Phi^* \Pi.
\end{equation}
The generators for the noise have been chosen to be the gauge-invariant observable quantities $\{E_i,|\Phi|^2\}$, in line with previous work on superconducting films that made use of the Krasnitz method\cite{Stephens:2001fv,Bettencourt:2003qb}. This physically-motivated choice is not unique.

Defining covariant derivatives
\begin{equation}
D^+_i \Phi(x) = \frac{U_i(x)\Phi(x+a\hat{\imath}) - \Phi(x)}{a}; \qquad D^-_i \Phi(x) = \frac{\Phi(x) - U_i^*(x) \Phi(x-a\hat{\imath})}{a},
\end{equation}
the stochastic field equations are then
\begin{eqnarray}
\dot{A}_i &=& E_i + \left(\beta
\dot{E}_i + \Gamma\right) \\
\dot{E}_i &=& \sum_{jklm} \epsilon_{ijk}
\epsilon_{klm} \Delta_j^- \Delta_l^+ A_m(x) - \frac{2e}{a} \sum_{x,i}
\mathrm{Im} \, \Phi^* (x) D_i^+ \Phi(x) \\
\dot{\Phi} &=& \Pi \\
\dot{\Pi} &=&  \sum_i D_i^{-} D_i^{+} \Phi (x)
- m^2 \Phi^* -  \lambda  (\Phi^* \Phi) \Phi - \Phi
\left( \beta^\Pi  \partial_t |\Phi|^2 + \Gamma^\Pi \right)
\end{eqnarray}
where $\Gamma$ and $\Gamma^\Pi$ are the Gaussian noise terms for the gauge and scalar fields, respectively, satisfying fluctuation-dissipation relations
\begin{eqnarray}
\left< \Gamma(x,t)\Gamma(x',t') \right> & = &
2 \beta T \delta^{(3)}(x - x') \delta(t-t') \\
\left< \Gamma^\Pi(x,t)\Gamma^\Pi(x',t') \right> & =
& 2 \beta^{\Pi} T \delta^{(2)}(x - x') \delta(t-t').
\end{eqnarray}
The dimensionalities of the delta functions differ due to the scalar
field lying in a plane. We will take $\beta=0.5$, $\beta^{\Pi} = 0.25$, $T=0.01$, $m^2=1$, $e=0.1$ and $\lambda=0.1$; this places us in the type II regime with $\kappa = 2\sqrt{\lambda}/e \approx 6.3$. We have checked that our results do not depend on lattice spacing, but for convenience we generally work with $a=0.5$. The values of $\beta$ and $\beta^{\Pi}$ were chosen such that the system behaves in a heavily damped manner after the quench.

\subsection{Boundary conditions and External field}
Periodic boundary conditions are used at all times. This forces the total magnetic flux through each of the boundaries to be zero. Where we wish to impose an external field, we do so by twisting a single plaquette located at an arbitrary position $x_0$ in the $x-y$ ($1-2$) plane of the gauge part of the Hamiltonian\cite{Kajantie:1998zn}. The effect of this is to add the term
\begin{equation}
H_\mathrm{ext} = a\sum_{x} \left(A_1(x) + A_2(x+a\hat{1}) - A_1(x+a\hat{2}) - A_2(x)\right)\phi_\mathrm{ext} \,\delta^{(2)} (x - x_0)
\end{equation}
to the Hamiltonian, where $\phi_\mathrm{ext}$ is the applied flux. We are free to make this greater than $2\pi$, because we have used the noncompact formulation of the gauge field. The equations of motion must of course be modified appropriately. It is possible to think of this external field as a Dirac string running through one plaquette on each slice in the $z$-direction.

\subsection{Quench protocol}
The stochastic equations given above are used to simulate the system throughout (although the same results could be obtained at less computational cost by using an initial thermal ensemble obtained from a Monte Carlo simulation and then evolving with overdamped dynamics). Given an initial `cold' system, the system was thermalised until the magnetic flux density in the $z$ direction was constant throughout the lattice and the time-averaged total energy was no longer varying. A linear quench was then carried out to take the system into the superconducting phase. Starting from $t=-\tau_\mathrm{Q}$, $m^2$ was changed from its initial value as follows
\begin{equation}
m^2(t) =\left\{\begin{array}{cl}
-m_0^2 (t/\tau_\mathrm{Q}),\qquad  & -\tau_\mathrm{Q}<t<\tau_\mathrm{Q} \\
-m_0^2, \qquad & t>\tau_\mathrm{Q}
   \end{array}\right.
\end{equation}
At $t=8\tau_\mathrm{Q}$ the simulation was stopped and the winding number $n$ was measured; the total magnetic flux trapped in the ring was also measured (as a cross check).

\section{Results}

The system was simulated as described in Section~\ref{sec:simulations}. Each data point is the result of 600-1000 individually thermalised runs; the results presented in this paper represent about five thousand hours of computer time. Error estimates were obtained from a bootstrap resampling of the results. We verified that the results did not change when the border around the film was enlarged, nor when $L_z$ was increased.

\subsection{Finite size}

\begin{figure}
\begin{center}
\includegraphics[width=0.65\linewidth,keepaspectratio,angle=270]{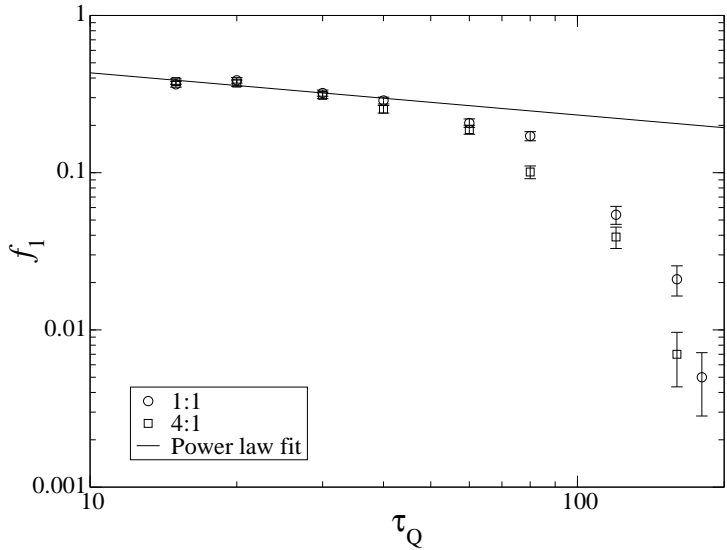}
\end{center}
\caption{Probability of observing one defect, $f_1$, as a function of quench time $\tau_\mathrm{Q}$. Results are shown for two aspect ratios. As observed in the one-dimensional simulations, a change in behaviour from KZ scaling to exponential suppression occurs. The power law fit shown is to the first four points only and yields an exponent $\sigma = 0.26\pm0.05$. Incorporating additional points produced a poorer fit; we associate these with the exponential damping regime discussed in the text.}
\label{fig:tq}
\end{figure}

With $B_\mathrm{ext}=0$, we first sought to see if the exponential falloff in trapping probability predicted\cite{WeirRivers} by 1D simulations in Eq.~(\ref{eq:expdamping}) persisted for the more realistic theory. It does. In Fig.~\ref{fig:tq}, we show the crossover from KZ behaviour, indistinguishable from the falloff obtained previously\cite{WeirRivers}. Differing aspect ratios were tried for the central hole for the same perimeter, ranging from square to an elongated 4:1 rectangle, but no deviation from the perimeter law behaviour of Eq.~(\ref{eq:kzlaw}) was detected. In this parameter range, then, we expect that the KZ mechanism (and its extension to small volumes) will be the only detectable influence on fluxoid formation, with no discernible evidence for magnetic field correlations.

\subsection{External fields}

\begin{figure}
\begin{center}
\includegraphics[width=0.65\linewidth,keepaspectratio,angle=270]{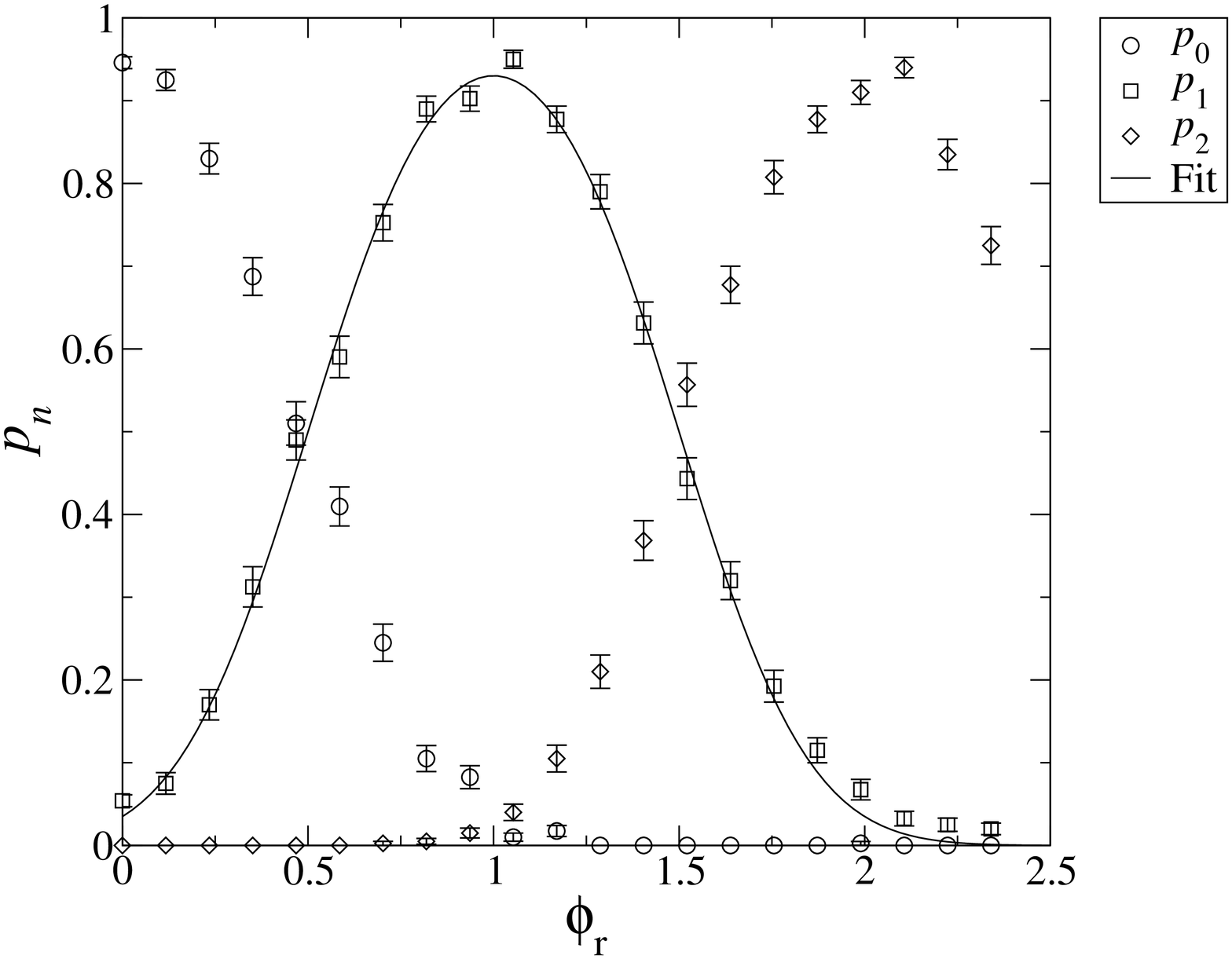}
\end{center}
\caption{Probability of observing zero ($p_0$), one ($p_1$) and two ($p_2$) defects as a function of normalised external magnetic flux $\phi_\mathrm{f}$, with $\tau_\mathrm{Q} = 120$. The curve is a one parameter fit of the form given in Eq.~(\ref{eq:f0}) for $p_1$, and yields $s=0.39 \pm 0.01$.}
\label{fig:ext}
\end{figure}

The results for nonzero external field are shown in Fig.~\ref{fig:ext}, with fits of the form in Eq.~(\ref{eq:f0}) also shown. Despite the simple assumptions going into this ansatz, it agrees with the results of the simulations very well indeed. We tested the ansatz at various $\tau_\mathrm{Q}$ and found that it worked well throughout the range shown in Fig.~\ref{fig:tq}.

\section{Discussion}

We have simulated the effect of a temperature quench on a two-dimensional superconductor with its central area removed, embedded in three-dimensional space (see Fig.~\ref{fig:setup}) as it is driven through its phase transition. The results of our simulations are two-fold.

Firstly, in the absence of an external magnetic bias field, for the trapping rates studied there is spontaneous production of fluxoids which, for rapid quenches, accords with the KZ scaling behaviour of Eq.~(\ref{eq:kzlaw}) and which, for slow quenches, shows exponential falloff; there is no perceptible influence of the magnetic field freezing out. Recent experiments have also struggled to show this effect\cite{Golubchik}. Further, the fluxoid density depends only on the inner perimeter and not on the aspect ratio of the superconducting rectangle. In all of these regards, there is no difference between the full theory and what we have observed in an idealised one-dimensional ring\cite{WeirRivers} with Gaussian winding number density.

Secondly, experimentally it is difficult to eliminate stray magnetic fields when carrying out temperature quenches of superconducting rings. Such fields artificially increase the likelihood of seeing fluxoids. Although the simulation of Fig.~\ref{fig:ext} is for zero stray residual field $\phi_\mathrm{r}$ we now know how to proceed to eliminate their possible effects. We apply external fields and identify the minimum (true) value of $f_1$ by matching the profiles to those of Fig.~\ref{fig:ext}. Again, the full simulation is well represented by a theoretical Gaussian ansatz that models the effect of an external (non-dynamical) field on fluxoid production.

The relevance of this analysis is that experiments are being prepared to provide a rigorous test of the simple KZ scenario for superconducting loops, originally proposed by Zurek more than 25 years ago but, as yet, only partially attempted\cite{2009PRB}.

\begin{acknowledgements}
DJW would like to thank Arttu Rajantie and Anders Tranberg for useful
discussions. The simulations in this paper were performed using the Imperial College High Performance Computing Service.
\end{acknowledgements}

\pagebreak


\begin{thebibliography}{10}

\bibitem{Kibble:1976sj}
T. Kibble, \textsl{J. Phys. A} \textbf{9}, 1387, (1976)

\bibitem{Kibble:1980mv}
T. Kibble, \textsl{Phys. Rept.} \textbf{67}, 183, (1980)

\bibitem{Zurek:1985qw}
W. Zurek, \textsl{Nature} \textbf{317}, 505, (1985)

\bibitem{Zurek:1993ek}
W. Zurek, \textsl{Acta Phys. Polon. B} \textbf{24}, 1301, (1993)

\bibitem{Polturak}
D. Golubchik, E. Polturak, and G. Koren, \textsl{Phys. Rev. Lett.} \textbf{104}, 247002, (2010)

\bibitem{2006PRL}
R. Monaco, J. Mygind, M. Aaroe, R. Rivers, and V. Koshelets, \textsl{Phys. Rev. Lett.} \textbf{96}, 180604. (2006)

\bibitem{2009PRB}
R. Monaco, J. Mygind, R. Rivers, and V. Koshelets, \textsl{Phys. Rev. B} \textbf{80}, 180501, (2009)

\bibitem{Yates:1998kx}
A. Yates and W. Zurek, \textsl{Phys. Rev. Lett.} \textbf{80}, 5477, (1998)

\bibitem{Donaire:2004gp}
M. Donaire, T. Kibble, and A. Rajantie, \textsl{New J. Phys.} \textbf{9}, 148, (2007)

\bibitem{WeirRivers}
D. Weir and R. Rivers, \textsl{J. Phys.: Conf. Ser.} \textbf{286}, 012056, (2011)

\bibitem{Hindmarsh:2000kd}
M. Hindmarsh and A. Rajantie, \textsl{Phys. Rev. Lett.} \textbf{85}, 4660, (2000)

\bibitem{Krasnitz:1995xi}
A. Krasnitz, \textsl{Nucl. Phys. B} \textbf{455}, 320, (1995)

\bibitem{Stephens:2001fv}
G. Stephens, L.M. Bettencourt, and W. Zurek, \textsl{Phys. Rev. Lett.} \textbf{88}, 137004, (2002)

\bibitem{Bettencourt:2003qb}
L. Bettencourt and Stephens, G., \textsl{Phys. Rev. E} \textbf{67}, 066105, (2003)

\bibitem{Kajantie:1998zn}
K. Kajantie, M. Laine, T. Neuhaus, J. Peisa, A. Rajantie, and K. Rummukainen, \textsl{Nucl. Phys. B} \textbf{546}, 351, (1999)

\bibitem{Golubchik}
D. Golubchik, E. Polturak, G. Koren, B. Shapiro, and I. Shapiro, \textsl{J. Low Temp. Phys.} \textbf{164}, 74, (2011)











\end{thebibliography}
\end{document}